\documentclass[prx,superscriptaddress,amsmath,amssymb,floatfix,twocolumn,showpacs,amsfonts,longbibliography,prx]{revtex4-2}
\usepackage{times}
\usepackage[varg]{txfonts}
\usepackage{textcomp}
\usepackage{graphicx}
\usepackage{subfigure}
\usepackage{tabu}
\usepackage{color}
\usepackage[plainpages=false,pdfpagelabels,colorlinks=true,linkcolor=red,urlcolor=blue,citecolor=blue,pdftitle={Title},pdfauthor={},pdfdisplaydoctitle=true,pdfduplex=DuplexFlipLongEdge]{hyperref}
\usepackage{braket}
\usepackage{float}
\usepackage{overpic}
\usepackage{gensymb}
\usepackage{mathrsfs}
\usepackage{tikz}
\usepackage{bm}
\usepackage{multirow}
\usepackage{amsfonts}
\usepackage{amsmath}
\usepackage{mathrsfs}
\usepackage{amssymb}
\usepackage{paralist}
\usepackage{indentfirst}
\usepackage{ulem}
\usepackage{soul}
\usepackage{cancel}
\usepackage{CJK}

\allowdisplaybreaks

\hypersetup{
           breaklinks=true,   % splits links across lines
           colorlinks=true,   % displays links as colored text instead of blocks
           pdfusetitle=true,  % \title and \author values into pdf metadata
        }
        
\newcommand{\tcr}{\textcolor{red}}
\newcommand{\tcb}{\textcolor{blue}}

\begin{document}
\title{Spinon Singlet in Quantum Colored String: Origin of $d$-Wave Pairing in a Partially-Filled Stripe}
\author{Jia-Long Wang}
\affiliation{Beijing Computational Science Research Center, Beijing 100193, People's Republic of China}
\author{Shi-Jie Hu}
\email{Corresponding author: shijiehu@csrc.ac.cn}
\affiliation{Beijing Computational Science Research Center, Beijing 100193, People's Republic of China}
\affiliation{Department of Physics, Beijing Normal University, Beijing 100875, People's Republic of China}
\author{Xue-Feng Zhang}
\email{Corresponding author: zhangxf@cqu.edu.cn}
\affiliation{Center of Quantum Materials and Devices, Chongqing University, Chongqing 401331, China}
\affiliation{Department of Physics and Chongqing Key Laboratory for Strongly Coupled Physics, Chongqing University, Chongqing 401331, China}

\begin{abstract}
Although both experimental observations and numerical simulations have reached a consensus that the stripe phase is intertwined with superconductivity in cuprates, the microscopic mechanism behind $d$-wave pairing in the presence of stripes remains unclear.
Using the effective theory of quantum colored strings, we derive the wavefunction in Fock space.
Our results show that two spinons with opposite chiralities tend to pair into a spinon singlet, which in turn facilitates the formation of negative pair-pair correlations between distant $x$-bonds and $y$-bonds, a hallmark of the $d$-wave pairing pattern.
The same pair-pair correlation pattern is observed across various models, as confirmed by large-scale density matrix renormalization group calculations.
Based on these results, we conclude that the spinon singlet is the origin of $d$-wave superconductivity in a fluctuating, partially-filled stripe, and this mechanism may also extend to multi-stripe configurations.
%allowing us to identify each vector that contributes most significantly to the $d$-wave superconductivity correlations.
%Ultimately,
%The stripe phase consists of many fluctuating stripes that resemble numerous rivers flowing with electrons and holes.
%Recent numerical results demonstrate that long-range $d$-wave superconductivity correlations can be established within the stripe phase, thereby making the coexistence of stripe order and superconductivity possible.
\end{abstract}
\maketitle

%\section{Introduction}
%Introduction of the stripe order and the superconductivity
\textit{Introduction}.--Establishing the correct microscopic mechanisms for the pairing process is a crucial step in understanding high-$T_c$ superconductivity~\cite{BCS, hightc_review0, hightc_review1, hightc_review2, hightc_review3, hightc_review4}.
In hole-doped cuprates, experiments have provided strong support for the idea of local electron pairing within the widely observed $d$-wave superconducting states~\cite{paring_exp0, paring_exp1, paring_exp_yayu}.
Numerical simulation results, which are consistent with experimental detections~\cite{Exp_stripe1, Exp_stripe2}, have also suggested an intertwined relationship between the stripe phase and superconductivity~\cite{stripe_review1, stripe_review2}, extending beyond the Hubbard model~\cite{Hubbard_Science, Hubbard_PRL, Hubbard_PRX, Hubbard_shiwei, Hubbard_yuanyao} to include a wide range of models, e.g., $t$-$J$, $t$-$t’$-$J$, etc~\cite{White_stripe, t-J1, t-J_ext1, t-J_ext2, t-J_ext3, ttj_science, tj_jiang, ttj_liwei}.
More recently, simulations of the Hubbard model using density matrix renormalization group (DMRG) and quantum Monte Carlo (QMC) methods have revealed the coexistence of $d$-wave superconductivity with partially-filled stripes~\cite{Hubbard_shiwei}.
It was also found that Cooper pairs are not uniformly distributed in space, but tend to accumulate at the hole-rich stripes~\cite{Wietek_2022}, challenging Anderson's resonating valence bond (RVB) paradigm~\cite{anderson_rvb}.
However, the origin of the pseudogap appears to be closely related to the melting of the stripe phase~\cite{paring_pg,pg_science}.
Therefore, investigating the role of stripes in $d$-wave pairing becomes a particularly promising and valuable research orientation.

\begin{figure}[t!]
\centering
\includegraphics[width=0.99\linewidth]{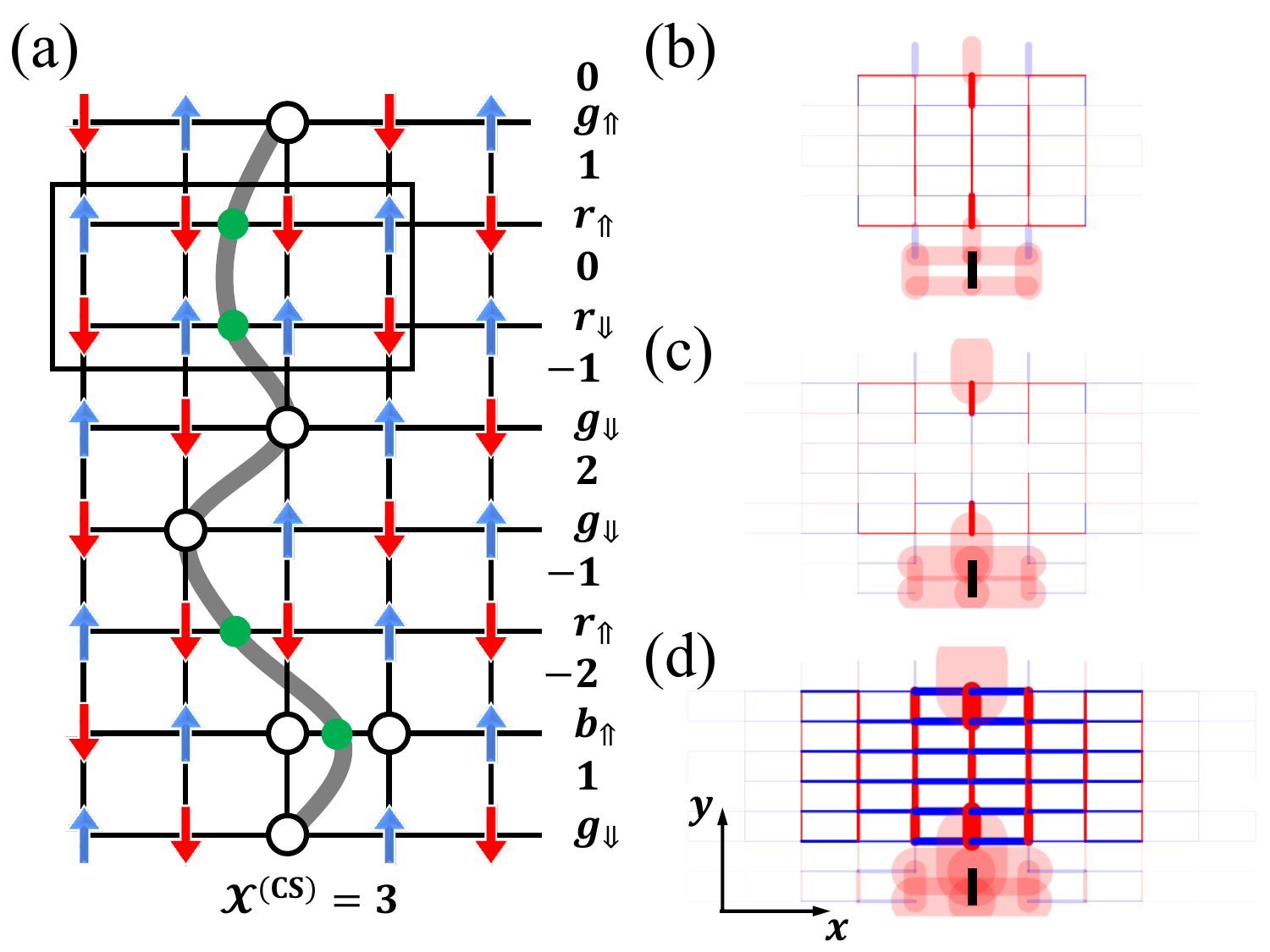}
\caption{(a) A schematic diagram shows the mapping from an electron configuration to a QCS (blue line) representation, with holes ($\circ$), spin-up (\tcb{$\uparrow$}), and spin-down (\tcr{$\downarrow$}) indicated.
The rectangle encloses a spinon singlet.
The PPC function, starting from the target $y$-bond (black line), is calculated using: (b) QCSM~\eqref{eq:effHam}, (c) DMRG in the $t$-$J_z$ model, and (d) DMRG in the $t$-$J$ model.
The width of the red and blue ribbons at each bond represents the magnitude of the positive and negative PPC, respectively.
To emphasize the long-range PPC in the central region, the ribbons at the bonds adjacent to the target $y$-bond are made transparent, with their widths halved.
In (b-d), $L_x=11$ and $L_y = 8$ are considered, where the ground state features a $3/4$ hole-doped stripe.
}\label{fig01}
\end{figure}

\begin{figure}[t!]
\centering
\includegraphics[width=0.99\linewidth]{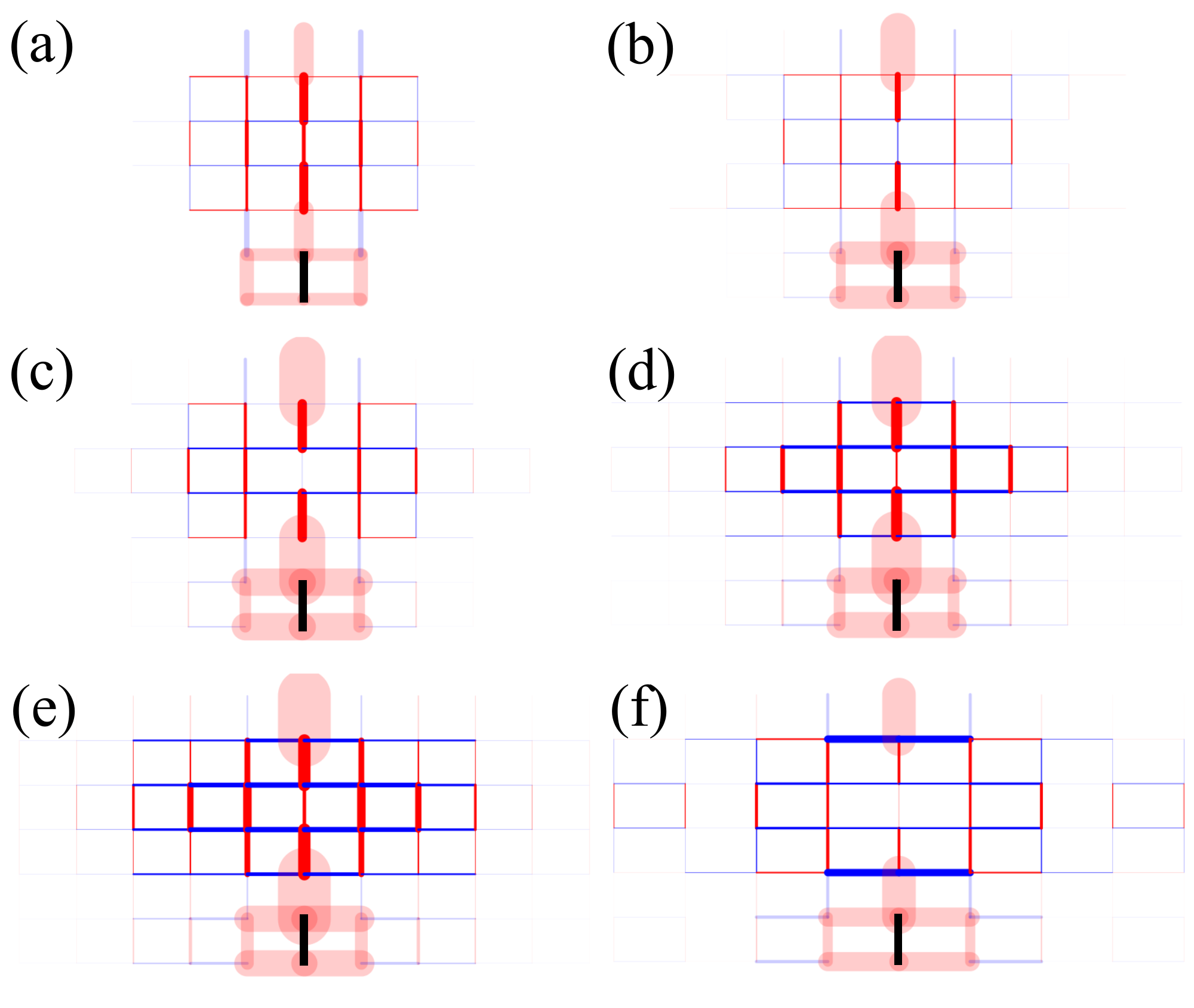}
\caption{The PPC function, starting from a target $y$-bond (black line), is calculated using: (a) QCSM~\eqref{eq:effHam}, (b-e) DMRG in the $t$-$J$-$\alpha$ model~\eqref{eq:extendedtjmodel} with (b) $\alpha=0$ ($t$-$J_z$ model), (c) $0.4$, (d) $0.8$ and (e) $1$ ($t$-$J$ model), and (f) DMRG in the Hubbard model including both NN and NNN hopping terms.
We fixed $L_y=6$ and chose $L_x=11$ in (a-e) and $9$ in (f), where the ground state has a $2/3$ hole-doped stripe.
}\label{fig02}
\end{figure}

%The quantum color string model
To describe the incommensurate spin and charge orders in the stripe phases, J. Zaanen introduced the concept of a quantum string with a $\pi$-phase shift, which arises from doping the antiferromagnetic (AF) spin ordering background~\cite{Zaanen_stripe, Zaanen01, string_Nishiyama}.
Recently, a quantum colored string (QCS) model (QCSM) has been proposed~\cite{color_string}, as shown in Fig.~\ref{fig01}(a).
In addition to Zaanen's original string model, the QCSM exhibits more features: (\textbf{1}) It is bottom-up derived from the $t$-$J_z$ model, thereby providing a quantitative rather than a phenomenological description. (\textbf{2}) The model includes hopping terms along the $y$-axis and additional fluctuations, allowing it to describe partially-filled stripes.
In this manuscript, we use the QCSM to demonstrate that a pair of spinons energetically forms a singlet when doped into a fully-filled stripe on a square lattice rolled into a cylinder.
The spinon singlet with the lowest local potential energy, as enclosed by the rectangle in Fig.~\ref{fig01}(a), plays a key role in facilitating $d$-wave pairing.
By analyzing the bases in the ground-state wavefunction that most significantly contribute to the pair-pair correlation (PPC) function $G_{b,b'} = \braket{\Delta^\dag_b \Delta_{b'}^{\phantom{\dag}}}$ between Cooper pairs at bonds $b$ and $b'$, we establish a microscopic scenario: QCS fluctuations, combined with the spinon singlet state, promote the emergence of $d$-wave pairing at long distances.
In large-scale numerical simulations using the state-of-the-art DMRG method, we also observe a clear and consistent $d$-wave pairing pattern across various models.
Specifically, $G_{b,b'}$ for a target $y$-bond and a distant $x$-bond exhibits a negative sign, as indicated by the blue ribbons in Figs.~\ref{fig01}(b-d) and \ref{fig02}.
Notably, the ground-state wavefunction of the QCSM closely matches that of the $t$-$J_z$ model as computed by DMRG.
Lastly, we examine a half-filled stripe on a larger cylinder with circumference $L_y = 8$ and find that the $d$-wave pairing remains stable.

In this work, we mainly study a $t$-$J$-$\alpha$ model, described by the following Hamiltonian:
\begin{eqnarray}\label{eq:extendedtjmodel}
	\begin{split}
	    H =&-t\sum_{\langle \ell,\ell'\rangle,\sigma}\left(f_{\ell,\sigma}^\dag f_{\ell',\sigma}^{\phantom{\dag}}+\textrm{h.c.}\right)\\
	       &+J\sum_{\langle \ell,\ell'\rangle} \left[ \alpha \left(S_\ell^x S_{\ell'}^x + S_\ell^y S_{\ell'}^y \right) + S_\ell^z S_{\ell'}^z - \frac{1}{4} n_\ell^{\phantom{\dag}} n_{\ell'}^{\phantom{\dag}} \right]\, ,
    \end{split}
\end{eqnarray}
where $f_{\ell,\sigma}^\dag$, $f_{\ell,\sigma}^{\phantom{\dag}}$ and $n_{\ell,\sigma}^{\phantom{\dag}} = f_{\ell,\sigma}^\dag f_{\ell,\sigma}^{\phantom{\dag}}$ represent the creation, annihilation, and particle number operators for an electron with spin $\sigma=\uparrow$,$\downarrow$ at site-$\ell$, respectively.
The term $\Delta_b = \left(f_{j,\downarrow} f_{i,\uparrow} - f_{j,\uparrow} f_{i,\downarrow}\right)$ describes the annihilation of a Cooper pair at bond $b = (i,\ j)$, where site $i$ has a higher index than site $j$, according to the early site-labeling convention~\cite{color_string}.
The same applies to the bond $b'=(i',\ j')$.
The particle number operator at site-$\ell$ is defined as $n_\ell = n_{\ell,\uparrow} + n_{\ell,\downarrow}$, while the spin components at site-$\ell$ are given by $S^x_\ell$, $S^y_\ell$ and $S^z_\ell$.
The nearest-neighboring (NN) hopping amplitude $t=1$ defines the energy unit, and the anisotropy parameter $\alpha \in [0,\ 1]$ is tunable.
We set the spin interaction strength to $J=0.6$, which favors a $\pi$-phase stripe in the ground state.
To prevent the formation of a frozen string, we double the value of $J$ along the bonds connecting sites in the two edge columns~\cite{color_string}.
For the Hubbard model~\cite{Hubbard_shiwei}, we set the on-site repulsion strength to $U=8$ and introduce a next-nearest-neighboring (NNN) hopping amplitude of $t'=-0.2$ to stabilize the partially-filled stripes.
In both the $t$-$J$-$\alpha$ and Hubbard models, we use the U($1$) and SU($2$) DMRG methods to obtain the ground-state wavefunction $\ket{\psi_\text{D}}$, respectively.
Unless otherwise specified, a truncation dimension of $M=8192$ is used, ensuring truncation errors remain below $3 \times 10^{-4}$.
For consistency in the following discussions, we always choose odd values for $L_x$.
While selecting other values of $L_x$ or using irregularly sheared open boundaries may introduce certain phases to the spinon singlet, but it does not alter conclusions regarding $d$-wave pairing.

%\section{Quantum Colored String}\label{sec2}
%Quantum Color string
\textit{Effective theory}.--The Nagaoka mechanism dictates that a single hole remains confined within the AF background~\cite{nagaoka01, nagaoka02}.
However, when multiple holes are introduced, they can spontaneously organize into a stripe with a $\pi$-phase shift.
This one-dimensional, curved structure is quantum in nature, rather than classical, due to electron hopping, and is hence referred to as a ``quantum string''~\cite{string_peter, dqcp01, zhou01, wanyuan, supersolid}.
We propose a QCS composed of three types of color quasi-particles (CQPs): spinons $\uparrow\downarrow\downarrow\uparrow$ or $\downarrow\uparrow\uparrow\downarrow$, holons $\uparrow\circ\downarrow$ or $\downarrow\circ\uparrow$, and dual-holes $\uparrow\circ\circ\uparrow$ or $\downarrow\circ\circ\downarrow$, each labeled with color indices $c = \textbf{r}$, $\textbf{g}$ and $\textbf{b}$.
CQPs are further distinguished by their chirality $\chi$, defined according to the orientation of the leftmost spin.
For example, for spinons, we have $\textbf{r}_\Uparrow=\uparrow\downarrow\downarrow\uparrow$ and $\textbf{r}_\Downarrow=\downarrow\uparrow\uparrow\downarrow$.
To describe the displacement of CQPs along the $x$-axis between adjacent rows, we introduce an effective spin field (ESF) $\Gamma^z_{\bar{y}} \in \mathbb{Z}$, where $\bar{y}$ labels the dual row in between rows $y$ and $y+1$.
Without loss of generality, we choose the position of the CQP in row $y=1$ to label the position of QCS $\mathcal{X}^{(\text{CS})}$.
As shown in Fig.~\ref{fig01}(a), the electron configuration has a QCS representation: $\ket{s} = \ket{\mathcal{X}^{(\text{CS})}, \{\Gamma^z\}, \{c_\chi\}}$, where the sequences $\{\Gamma^z\}$ and $\{c_\chi\}$ represent the ESFs and CQPs across all rows, respectively. The Hilbert space is truncated to $\Omega$ due to specific constraints and conservation laws.

%The corresponding creation and annihilation operators are denoted as $c_{\chi}^\dag$ and $c_{\chi}$, respectively.

%The interaction term of the interaction of the CS
Building on the $t$-$J_z$ model~\cite{color_string}, the effective model for a QCS can be explicitly derived as:
\begin{eqnarray}\label{eq:effHam}
H^\text{CS}_\text{e} = H_\text{d} + H_\text{o}^{(\textbf{g})} + H^{(\textbf{r}\text{-}\textbf{b} / \textbf{g}\text{-}\textbf{g})}_\text{o} + H^{(\textbf{r}\text{-}\textbf{g}/\textbf{g}\text{-}\textbf{r})}_\text{o} + H^{(\textbf{b}\text{-}\textbf{g}/\textbf{g}\text{-}\textbf{b})}_\text{o}\, .
\end{eqnarray}
The diagonal term $H_\text{d}$, proportional to $|\Gamma^z|$, represents an attractive potential between CQPs.
The off-diagonal terms describe various interaction processes between CQPs: $H_\text{o}^{(\textbf{g})}$ accounts for electron hopping along the $x$-axis, while $H^{(\textbf{r}\text{-}\textbf{b} / \textbf{g}\text{-}\textbf{g})}_\text{o}$ reflects the creation of a \textbf{r}-\textbf{b} pair and its reverse process, and $H^{(\textbf{r}\text{-}\textbf{g}/\textbf{g}\text{-}\textbf{r})}_\text{o}$ and $H^{(\textbf{b}\text{-}\textbf{g}/\textbf{g}\text{-}\textbf{b})}_\text{o}$ describe the movement of spinons and dual-holes, respectively.
Clearly, the diagonal term $H_\text{d}$ suppresses the vibration of the QCS, while the quantum fluctuations introduced by the off-diagonal terms enhance this vibration.
The ground-state wavefunction $\ket{\psi_\text{E}} = \sum_{s \in\Omega} \alpha_s^{\text{E}} \ket{s}$ of the effective QCSM~\eqref{eq:effHam} is obtained through exact diagonalization (ED), where $\alpha_s^{\text{E}}$ gives the expansion coefficient of the basis $\ket{s}$.
Systematic benchmarks of this model have been presented in Ref.~\cite{color_string}.

\begin{figure}[t!]
\centering
\includegraphics[width=0.99\linewidth]{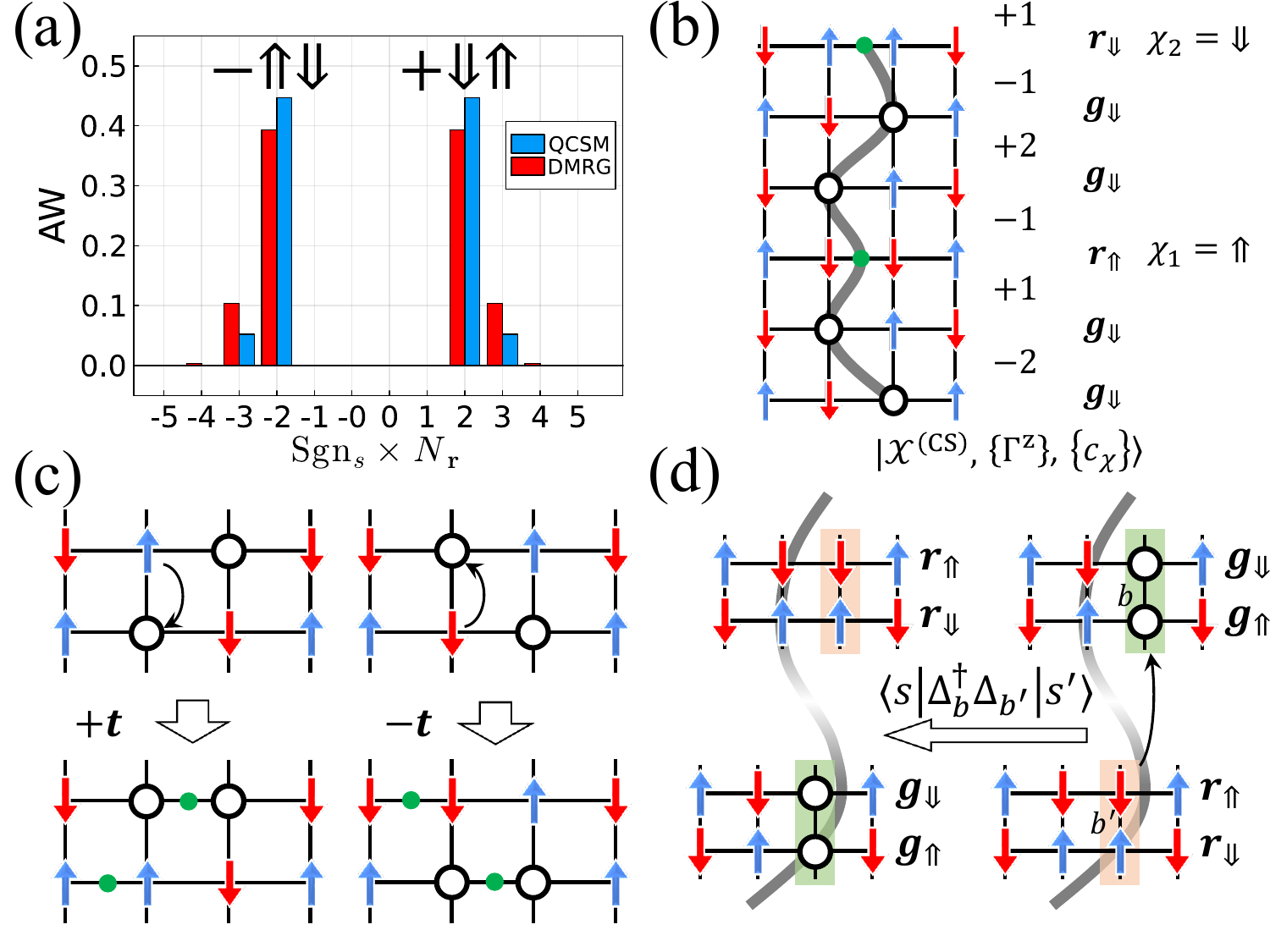}
\caption{(a) Distributions of accumulated weights (AW) in $\ket{\psi_\text{E}}$ and $\ket{\psi^\Omega_\text{D}}$ for a $2/3$ hole-doped stripe in the $t$-$J_z$ model with $L_x = 11$, $L_y = 6$ and $J_z = 0.6$.
The bases are labeled by $\text{Sgn}_s$ and $N_\textbf{r}$.
Schematic pictures for (b) defining the chirality sequence, (c) local vacuum fluctuation processes, and (d) the processes of moving a Cooper pair from $y$-bonds $b'$ to $b$.
}\label{fig03}
\end{figure}

%Terminology
\textit{Two-Spinon QCS}.--When a fully-filled stripe is doped with two electrons, or equivalently two additional spinons, it forms ``two-spinon QCS", which is a special case of partially-filled stripes.
For $L_y=6$ and $|\Gamma^z| \leq 5$ in the effective model~\eqref{eq:effHam}, the Hilbert space dimension reaches $286,098$, which is about $1\%$ of the dimension used in the DMRG calculations.
After projecting the wavefunction $\ket{\psi_\text{D}}$ onto the space $\Omega$, we obtain the wavefunction $\ket{\psi_\text{D}^\Omega} = \sum_{s \in \Omega} \alpha^\text{D}_s \ket{s}$, where $\alpha^\text{D}_s = \braket{s | \psi_\text{D}}$.
We then calculate the renormalized fidelity $\mathcal{F}_\text{R} = \lvert \braket{\psi_\mathrm{E} | \psi_\mathrm{D}^\Omega} \rvert / W$, with $W = \braket{\psi_\text{D}^\Omega | \psi_\text{D}^\Omega}$ representing the weight.
Using $M=2048$ in DMRG, we find $\mathcal{F}_\text{R} \approx 96.0\%$ with $W \approx 67.8\%$ for $J=1$, and $\mathcal{F}_\text{R} \approx 90.8\%$ with $W \approx 44.3\%$ for $J=0.6$ (see App.~B for details).
These results indicate that the wavefunction $\ket{\psi_\text{E}}$ contains the key information about $\ket{\psi_\text{D}}$.
As illustrated in Figs.~\ref{fig01}(b-d) and \ref{fig02}(a-e), the two-spinon QCS exhibits a $d$-wave pairing pattern in the $t$-$J$-$\alpha$ model, with $\alpha$ increasing smoothly.
In Table.~\ref{tabel1}, the fidelity values stay above $0.8$ throughout the process, except near $\alpha=1$, where the emergence of SU($2$) symmetry causes a reshaping of weight distribution in $\ket{\psi_\text{D}}$.
The $d$-wave pairing pattern is also observed in the Hubbard model, as seen in Fig.~\ref{fig02}(f).
%Therefore, it is reasonable to conclude that the $d$-wave pairing mechanism is embedded within $\ket{\psi_\text{E}}$ for the QCS.
Therefore, we can promisingly summarize the $d$-wave pairing mechanism from $\ket{\psi_\text{E}}$ for the QCS.

\begin{table}[h]
    \centering
    \begin{tabular}{|c|c|c|c|c|c|}
    \hline
     $\alpha$ & $0$ & $0.2$ & $0.4$ & $0.6$ & $0.8$ \\
     \hline
     $\mathcal{F} = \left\lvert \braket{\psi_\text{D} (\alpha + \delta\alpha) | \psi_\text{D} (\alpha)} \right\rvert$ & $0.903$ & $0.900$ & $0.886$ & $0.828$ & $0.438$ \\
      \hline
    \end{tabular}
    \caption{The fidelity $\mathcal{F}$ as a function of $\alpha$ in the $t$-$J$-$\alpha$ model is calculated with fixed $\delta\alpha=0.2$. We set $L_x = 11$ and $L_y=6$, ensuring that the ground state has a $2/3$ hole-doped stripe.}
    \label{tabel1}
\end{table}

%Wavefunction
The basis $\ket{s}$ can be characterized by the number of spinons $N_\textbf{r}$ and the sign $\text{Sgn}_s \equiv \text{sgn}(\alpha_s^{\text{E}})$.
As shown in Fig.~\ref{fig03}(a), the two-spinon bases with $N_\textbf{r}$=2 dominate the ground-state wavefunction, with the accumulated weights (AW) being symmetric with respect to $\text{Sgn}_s$=$\pm 1$.
The distribution of three-spinon bases exhibits similar features, though with much lower weights, allowing us to neglect the contributions from other bases with $N_\textbf{r}>3$.
Given that the lower and upper spinons in the two-spinon bases carry chiralities $\chi_1$ and $\chi_2$, we only need to consider configurations where $\chi_1 \ne \chi_2$, as the ground state lacks net magnetization.
Interestingly, we find that $\text{Sgn}_s$ for a two-spinon basis depends on the order of the chiralities, as highlighted in Fig.~\ref{fig03}(a).
Specifically, $\text{Sgn}_s = \mp 1$ for $\chi_1\chi_2 = \Uparrow \Downarrow$ and $\Downarrow \Uparrow$, respectively, as seen in the QCS configuration shown in Fig.~\ref{fig03}(b).
At leading order, the two-spinon QCS can be interpreted as a singlet pair of spinons moving along a fluctuating QCS.
The three-spinon bases $\ket{s}$ arise from local vacuum fluctuations of the two-spinon bases $\ket{s'}$, as illustrated in Fig.~\ref{fig03}(c).
If the emergent dual-hole is far from the two spinons in $\ket{s}$, the sign of its expansion coefficient is simply given by $\text{Sgn}_s = \text{Sgn}_{s'} \text{Sgn}_0$.
Here, the sign $\text{Sgn}_0$ depends on the specifics of electron hopping in the process from $\textbf{g}\text{-}\textbf{g}$ to $\textbf{r}\text{-}\textbf{b}$.
For instance, we find that $\text{Sgn}_0 = \mp 1$ for the left and right panels in Fig.~\ref{fig03}(c).
%In this work, we focus on the long-range correlation, which exhibits a clear $d$-wave pairing pattern.
Moreover, we examine the two-spinon QCS in a larger cylinder with $L_x=11$, $L_y=8$, and having a $3/4$ hole-doped stripe, where the two-spinon and three-spinon bases contribute $79.9\%$ and $19.0\%$ of the weight in $\ket{\psi_\text{E}}$, respectively.

\begin{figure}[t!]
\centering
\includegraphics[width=0.99\linewidth]{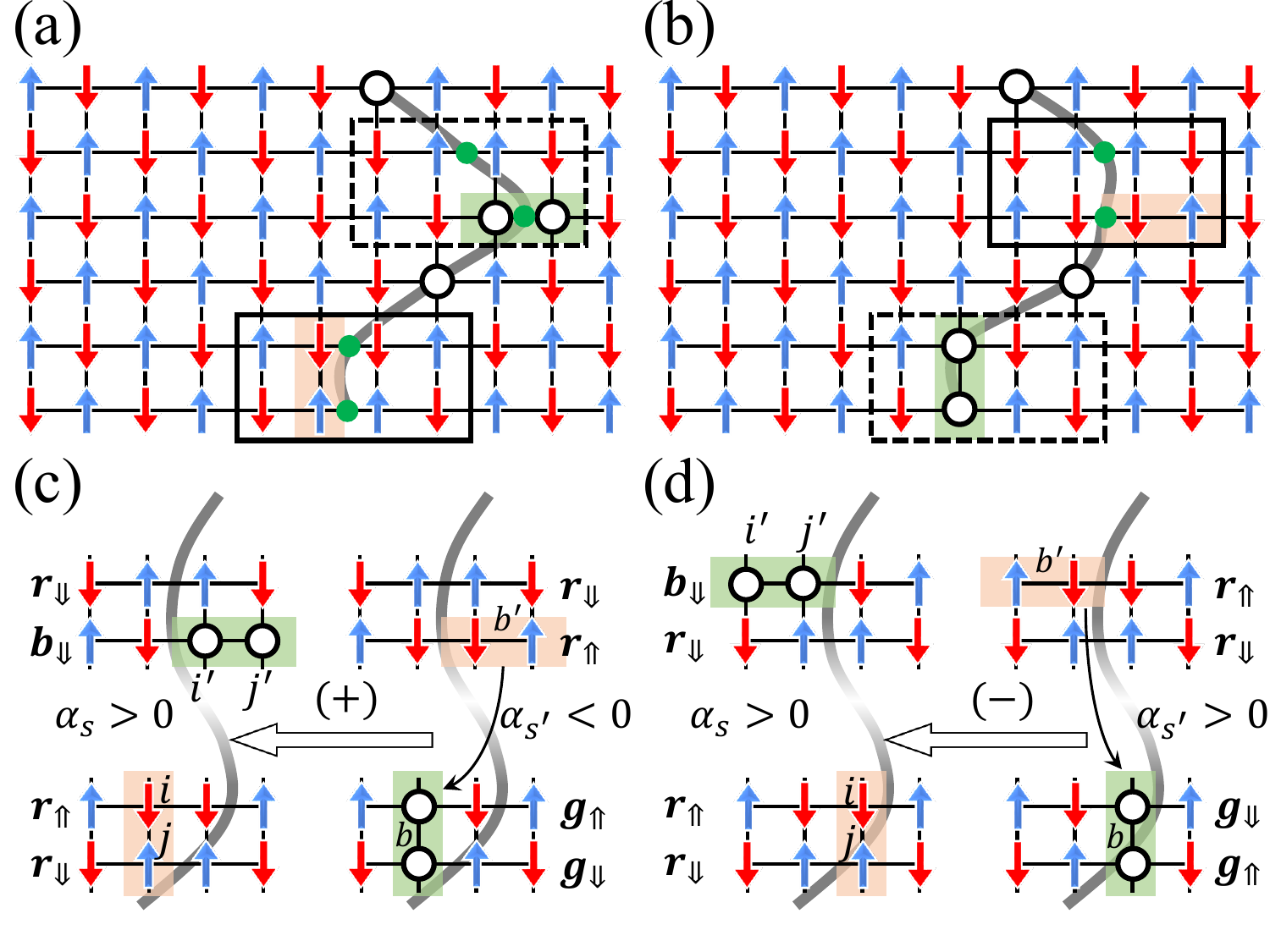}
\caption{Schematic diagrams illustrate a pair of (a) three-spinon and (b) two-spinon bases, which most significantly contribute to $G_{b,b'}$ for an $x$-bond $b'$ and a $y$-bond $b$ in a $2/3$ hole-doped stripe with $L_x = 11$ and $L_y=6$.
The dashed rectangles enclose a dule-hole-spinon in (a) and a hole-hole pair in (b), respectively.
The processes of moving a Cooper pair from $x$-bond $b'$ to $y$-bond $b$ (c) with and (d) without changing the chirality sequence are shown.
The sign of $\text{Sgn}_\Delta = \mp$ is indicated above the black hollow arrows, corresponding to cases (c) without and (d) with spin exchange, respectively.
}\label{fig04}
\end{figure}

%y-y bonds
\textit{$d$-wave pairing}.--Now, let us examine the PPC function between two $y$-bonds, $b$ and $b'$.
At leading order, as shown in Fig.~\ref{fig03}(d), the PPC function receives a contribution $G^{s,s'}_{b,b'} = \alpha^*_s \alpha^{\phantom{\dag}}_{s'} \braket{s | \Delta^\dag_b \Delta^{\phantom{\dag}}_{b'} | s'}$ from the two-spinon bases $\ket{s}$ (left panel) and $\ket{s'}$ (right panel), in which a pair of spinons always form a singlet state in two adjacent rows.
At long distances, the sign of $G^{s,s'}_{b,b'}$ is given by $\text{Sgn}\left(G^{s,s'}_{b,b'}\right) = \text{Sgn}_s \text{Sgn}_{s'} \text{Sgn}_\Delta$, where the sign $\text{Sgn}_\Delta$ accounts for the spin exchange between $\Delta_b$ and $\Delta_{b'}$.
In this setup, the operator $\Delta_{b'}$ annihilates two electrons with opposite spins at $y$-bond $b'$ in the basis $\ket{s'}$, and then $\Delta^\dag_{b}$ fills two holes at $y$-bond $b$ to generate the basis $\ket{s}$.
If the chirality sequence $\chi_1\chi_2$ in $\ket{s}$ and $\ket{s'}$ remains unchanged, then $G^{s,s'}_{b,b'} > 0$.
However, if the chirality sequence is altered, the sign $\text{Sgn}_\Delta = -1$ must be included, compensating for the sign change caused by the altered chirality sequence.
As a result, $G_{b,b'} > 0$ always for two $y$-bonds at long distances, owing to the spinon singlet state.
Similarly, $G_{b,b'}$ remains positive for two $x$-bonds, where a single spinon interacts with a dual-hole (see App.~A for details).

%y-bonds x-bonds
We also investigate the function $G_{b,b'}$ between a $y$-bond $b$ and an $x$-bond $b'$.
According to the effective theory, the nonzero contribution $G^{s,s'}_{b,b'}$ stems from a three-spinon basis $\ket{s}$ and a two-spinon basis $\ket{s'}$ at leading order.
The number of basis pairs that contribute nonzero $G^{s,s'}_{b,b'}$ is quite limited, especially for bonds at long distances.
This allows us to systematically identify all the contributing pairs and uncover the underlying microscopic pairing mechanism.

In Figs.~\ref{fig04}(a,b), one such pair is among the four basis pairs that together contribute over $86\%$ to $G_{b,b'}$ for the color-shaded bonds.
As shown in Fig.~\ref{fig04}(c), the three-spinon basis $\ket{s}$ originates from a two-spinon basis with a chirality sequence $\chi_1\chi_2 = \Downarrow \Uparrow$ and $\text{Sgn}_0 = +1$, giving $\text{Sgn}_s = +1$.
When the operators $\Delta_b^\dag$ and $\Delta^{\phantom{\dag}}_{b'}$ act on $\ket{s'}$, a three-spinon basis $\ket{s}$ is generated, involving no spin exchange, so $\text{Sgn}_\Delta=+1$.
Since the chirality sequence in $\ket{s'}$ changes, $\text{Sgn}_{s'} = -1$.
Notably, in $\ket{s'}$, two spinons form a singlet with $\Gamma^z = 0$.
Consequently, the contribution $G^{s,s'}_{b,b'}$ from this basis pair is negative, as $\text{Sgn}_s \text{Sgn}_{s'} \text{Sgn}_\Delta = -1$.
In contrast, another two-spinon basis $\ket{s''}$ (not shown), which can be generated with spin exchange, gives $\text{Sgn}_s=+1$ and $\text{Sgn}_{s''}=\text{Sgn}_\Delta=-1$, leading to a positive contribution $G^{s,s''}_{b,b'}$.
However, this contribution is smaller, as in the ground-state wavefunction $\ket{\psi_\text{E}}$ the weight for $\ket{s''}$ is lower compared to that for $\ket{s'}$.
This reduction results from the disruption of the spinon singlet in $\ket{s''}$, where $\vert \Gamma^z \vert \ne 0$, which increases the potential energy of the spinons.

%We can clearly observe that both vectors contain spinon pairs but with different orders of chirality sequences.

In Fig.~\ref{fig04}(d), we consider another pair of bases: a three-spinon basis $\ket{s}$ with $\text{Sgn}_s = +1$ and a two-spinon basis $\ket{s'}$ with $\text{Sgn}_{s'}=+1$.
In this case, the operators $\Delta_b^\dag$ and $\Delta_{b'}^{\phantom{\dag}}$ generate the basis $\ket{s'}$ with spin exchange, contributing a sign of $\text{Sgn}_\Delta = -1$.
After accounting for the positive contribution from the process without spin exchange, we find that the total contribution is negative.
In other instances, we also observe $G_{b,b'} < 0$ for $x$-bonds and $y$-bonds at long distances.
In sum, $G_{b,b'} < 0$ holds between an $x$-bond $b$ and a $y$-bond $b'$ at long distances.
We also present the data of $G_{b,b'}$ starting from an $x$-bond in App. A.

%Decaying of the amplitude
From the distribution of the PPC function, especially when the distance of two target bonds, $l_{b,b'}$, is large ($l_{b,b'} \gg 1$), as shown in Figs.~\ref{fig01}(b,c) and \ref{fig02}, we observe that its magnitude decreases as it moves away from the QCS along the $x$-axis.
This behavior can be intuitively understood, as CQPs produce a $d$-wave pairing pattern.
When the CQPs are farther apart, the large tension energy associated with the QCS strongly suppresses the weight of the corresponding bases.
Furthermore, when the spinon singlet is very close to the dual hole in space, their interference violet the simple behavior of $\text{Sgn}_0$, although the spinon singlet state still exists.
As a result, the $d$-wave pairing may weaken, or even be replaced by $s$-wave pairing~\cite{Chen_2024}.

\begin{figure}[t!]
\centering
\includegraphics[width=0.99\linewidth]{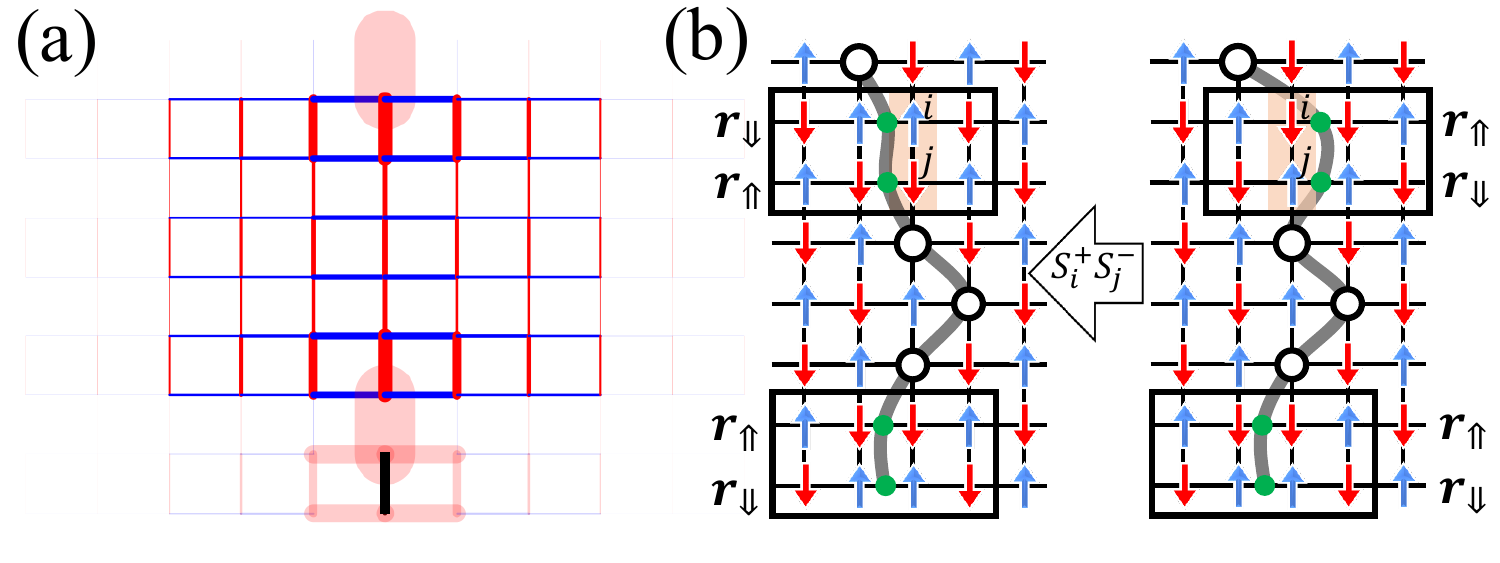}
\caption{(a) The PPC function $G_{b,b'}$ for a half-filled stripe in the $t$-$J$ model with $L_x=11$ and $L_y=8$, calculated using DMRG.
(b) A schematic diagram illustrates the movement of the spinon singlet (rectangle) resulting from a spin-flip operation.
}\label{fig05}
\end{figure}

\textit{Four-spinon QCS}.--For narrow cylinders with $L_y = 6$, the $\pi$-phase stripe is unstable because two neighboring holes tend to pair up, in agreement with the ``phase string" theory \cite{wenzhengyu}. However, when the circumference increases to $L_y = 8$, as shown in Fig.~\ref{fig05}(a), the DMRG results for the $t$-$J$ model ($\alpha=1$ in the model~\eqref{eq:extendedtjmodel}) demonstrate that a half-filled stripe with four doped holes becomes stable.
At the same time, the $d$-wave pairing pattern becomes distinctly robust.

%Spinon Paring state
In fact, the $d$-wave pairing pattern becomes stable as $\alpha$ increases.
Similarly, as shown in Figs.~\ref{fig02}(b-e), the spin AF $xy$-interaction significantly enhances the $d$-wave pairing, suggesting that the formation of the spinon singlet is also strengthened.
Although the effective theory has not yet fully captured the behavior of a partially-filled stripe within the $t$-$J$ model, we can still analyze the underlying physical mechanism.
As depicted in Fig.~\ref{fig05}(b), the spin-flipping operation on a $y$-bond moves the spinon singlet along the $x$-axis while simultaneously altering the chirality sequence.
Since the $xy$-interaction is of the AF type, these two spinons prefer to form a ``singlet" pair to lower energy.
Then, the ground state of a partially filled stripe with multiple spinons can be viewed as a Luttinger liquid of spinon singlets.
Of course, this conjecture requires further verification in future studies.

\textit{Conclusion and Outlook}.--Building on the quantum colored strings scenario originally proposed by us~\cite{color_string}, we have established a bottom-up effective theory to describe the correlations between distant Cooper pairs in the ground states of a partially-filled stripe.
In the $t$-$J_z$ model, we identify a clear microscopic mechanism in which the spinon singlet state, arising from a fluctuating two-spinon quantum colored string, leads to the emergency of $d$-wave pairing.
The spinon singlet is further stabilized by antiferromagnetic $xy$-interactions, which enhance the stability of the $d$-wave pairing, thus supporting our conclusions.
Based on the similar $d$-wave patterns observed in the pair-pair correlation functions of the $t$-$J$-$\alpha$ and Hubbard models, we conjecture that these models may share a common microscopic mechanism for pairing.

Our work provides a potential microscopic explanation for $d$-wave pairing in cuprates, especially when spin or charge stripes emerge, extending beyond Anderson's RVB paradigm.
Additionally, by placing two holons (acting as a vacuum in the dilute spionon case) above and below the spinon singlet, a 4$\times$4 structure with two holes can be formed.
This configuration matches the checkerboard plaquettes of size $4 a_0$ observed in underdoped Bi$_2$La$_x$Sr$_{2-x}$CuO$_{6+\delta}$ through STM~\cite{paring_exp_yayu}.
Furthermore, the formation of the pseudogap phase may be linked to vacuum bubble resulting from the breaking of the quantum colored string~\cite{Hubbard_yuanyao}.

%In the future work, we plan to analyse the interaction between spinon singlet in details. Meanwhile, the influence of three-site hopping \cite{tj_3hopping} should be also checked.

\section*{acknowledgment}
We would like to thank Hao Ding and Yuan-Yao He for many helpful discussions. This paper is also in memory of Prof. Peter Fulde and Prof. Jan Zaanen, who contributed a lot to the quantum strings. S. J. H. acknowledges funding from MOST Grant No. 2022YFA1402700, NSFC No. U2230402, NSFC Grant No. 12174020. X.-F. Z. acknowledges funding from the National Science Foundation of China under Grants No. 12274046, No. 11874094, No. 12147102, and No. 12347101, the Chongqing Natural Science Foundation under Grant No. CSTB2022NSCQ-JQX0018, the Fundamental Research Funds for the Central Universities Grant No. 2021CDJZYJH-003, and the Xiaomi Foundation/Xiaomi Young Talents Program.

%${\tikz[scale=0.8]{\draw plot [smooth, tension=1] coordinates { (0.,0.2) (-0.1,0.1) (0.0,0.0) (0.1,-0.1) (0,-0.2)};}}$

%\bibliographystyle{aipauth4-1}
\bibliography{ref}

\section*{appendix}
\subsection{PPC function starting from a target $x$-bond}
\begin{figure}[t!]
\centering
\includegraphics[width=0.99\linewidth]{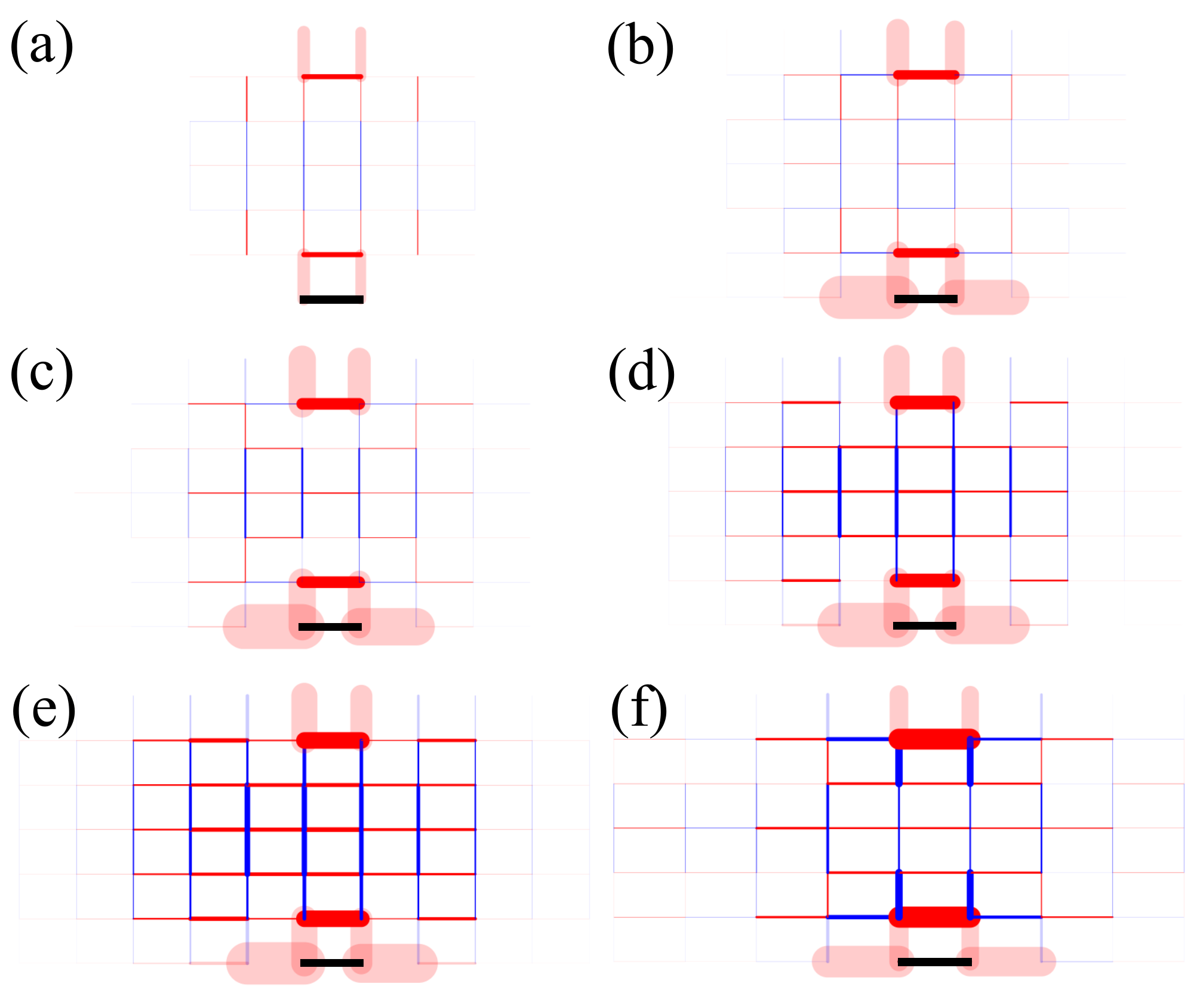}
\caption{The PPC function, starting from a target $x$-bond (black line), is calculated using: (a) QCSM~\eqref{eq:effHam}, (b-e) DMRG in the $t$-$J$-$\alpha$ model~\eqref{eq:extendedtjmodel} with (b) $\alpha=0$ ($t$-$J_z$ model), (c) $0.4$, (d) $0.8$ and (e) $1$ ($t$-$J$ model), and (f) DMRG in the Hubbard model.
Other parameters are the same as those in Fig.~\ref{fig02}.
}\label{figS1}
\end{figure}

\begin{figure}[h]
\centering
\includegraphics[width=0.99\linewidth]{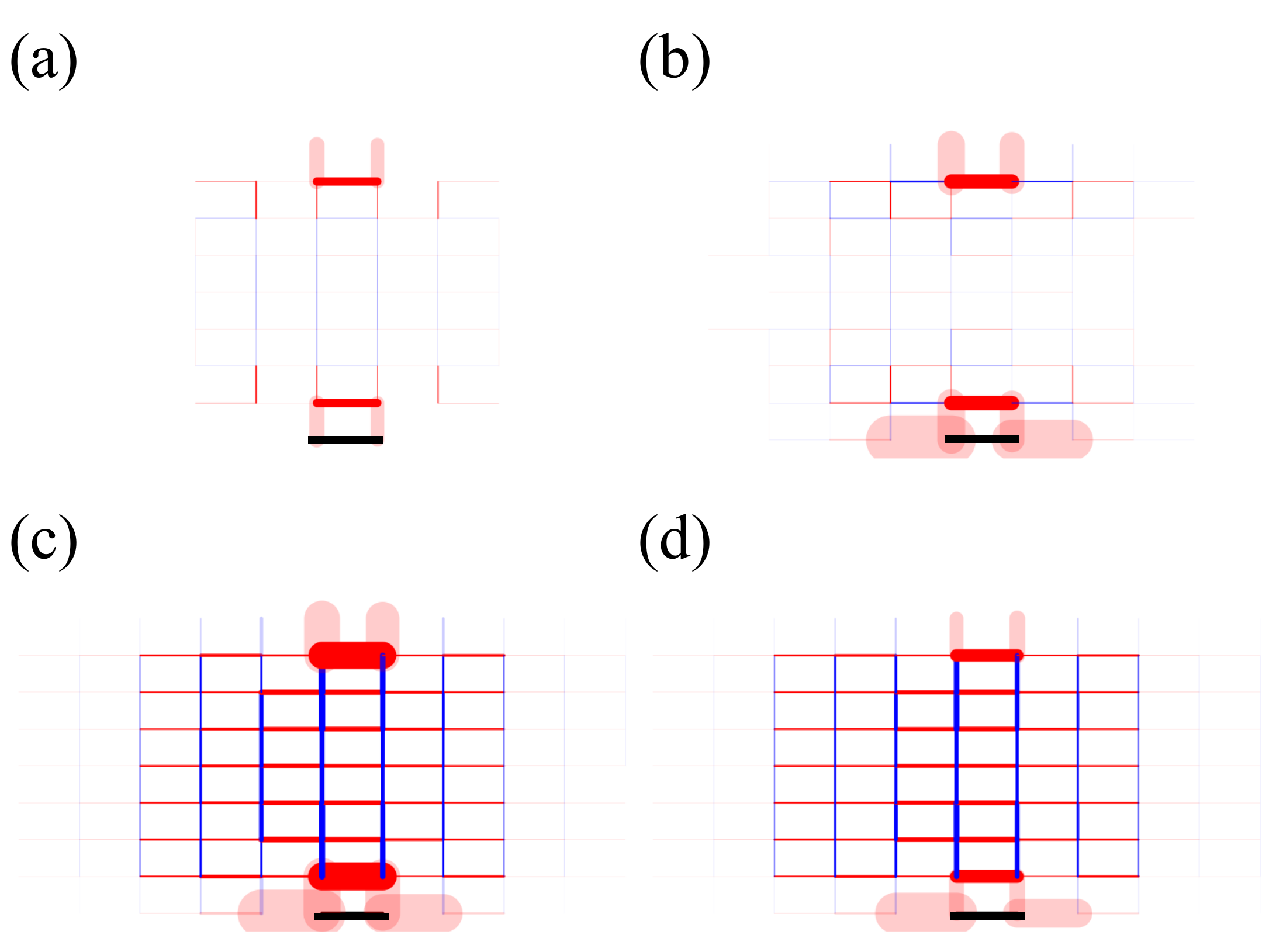}
\caption{The PPC function, starting from a target $x$-bond (black line), is calculated using: (a) QCSM~\eqref{eq:effHam}, and DMRG in (b) $t$-$J_z$ model, and (c,d) $t$-$J$ model.
The ground state has a $3/4$ hole-doped stripe in (a-c), while it has a $1/2$ hole-doped stripe in (d).
We use $L_x=11$ and $L_y=8$.
}\label{figS2}
\end{figure}

The $d$-wave pairing pattern means that the PPCs between $x$-bonds and $y$-bonds are negative, while those between $x$-bonds are positive.
As shown in Figs.~\ref{figS1}(a,b) for $L_y=6$, the correlations between a target $x$-bond and distant $y$-bonds in the $t$-$J_z$ model are found to be negative, as calculated using both QCSM and DMRG.
Additionally, these correlations are strongly enhanced while the $xy$ spin interactions are increased, and they remain stable in the Hubbard model, as demonstrated in Figs.~\ref{figS1}(c-f).
On the other hand, the correlations between $x$-bonds are very weak when $\alpha$ is small, but become positive and strong when $\alpha$ is enlarged.
Similar features can also be observed in a larger cylinder with $L_y=8$ demonstrated in Fig.~\ref{figS2}.

The corresponding mechanism can be comprehended using the effective theory established in the main text either.
Unlike other cases, the main contributions to the PCC between two $x$-bonds stem from the movement of a Cooper pair along the $x$-axis, involving two three-spinon bases, $\ket{s}$ and $\ket{s'}$, as depicted in Figs.~\ref{figS2}(e,f).
As the chirality sequence is altered when a spin exchange occurs between $\Delta_b$ and $\Delta_{b'}$, the sign of $G^{s,s'}_{b,b'}$ is always positive as long as the spinon-singlet is far away from the \textbf{r}-\textbf{b} pair.
Furthermore, the other processes related to the PCC between $x$-bonds can also be analyzed with similar procedures, and the overall conclusion remains unchanged.

\begin{figure}[t!]
\centering
\includegraphics[width=0.99\linewidth]{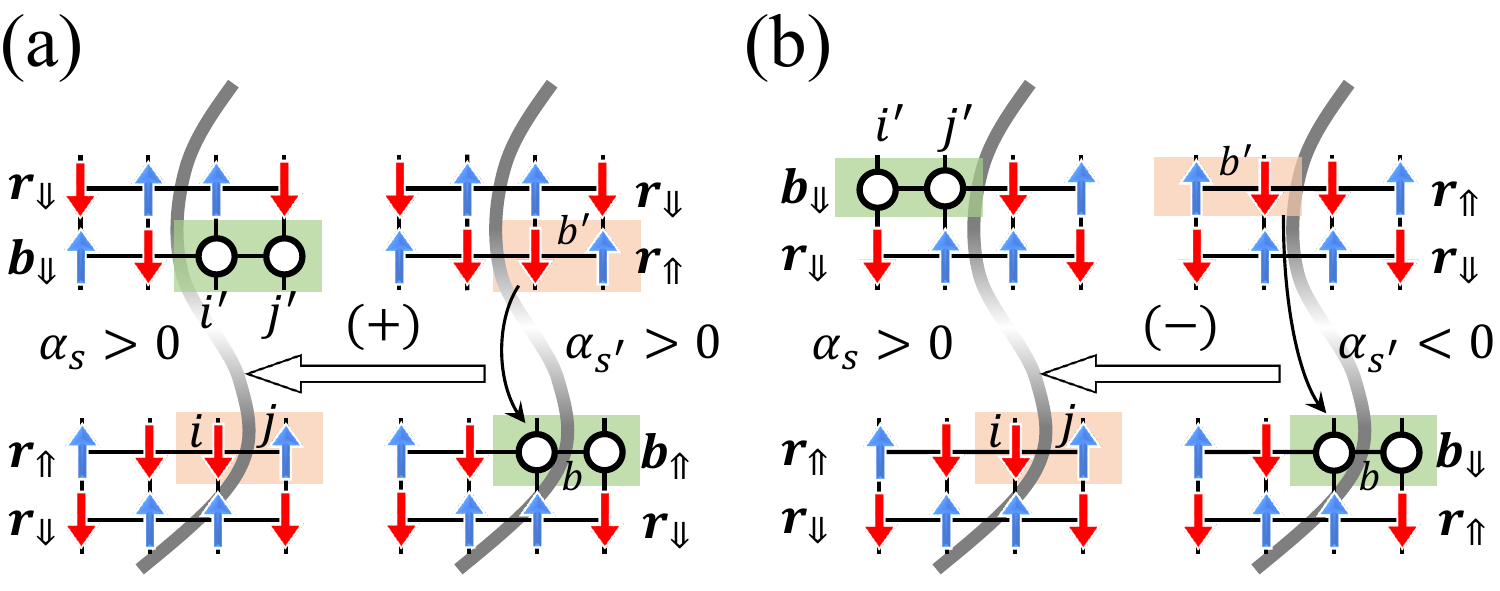}
\caption{Schematic diagrams illustrate the processes of moving a Cooper pair from $x$-bond $b'$ to $x$-bond $b$ (a) with and (b) without changing the chirality sequence.
}\label{figS3}
\end{figure}

\begin{figure}[b!]
\centering
\includegraphics[width=0.99\linewidth]{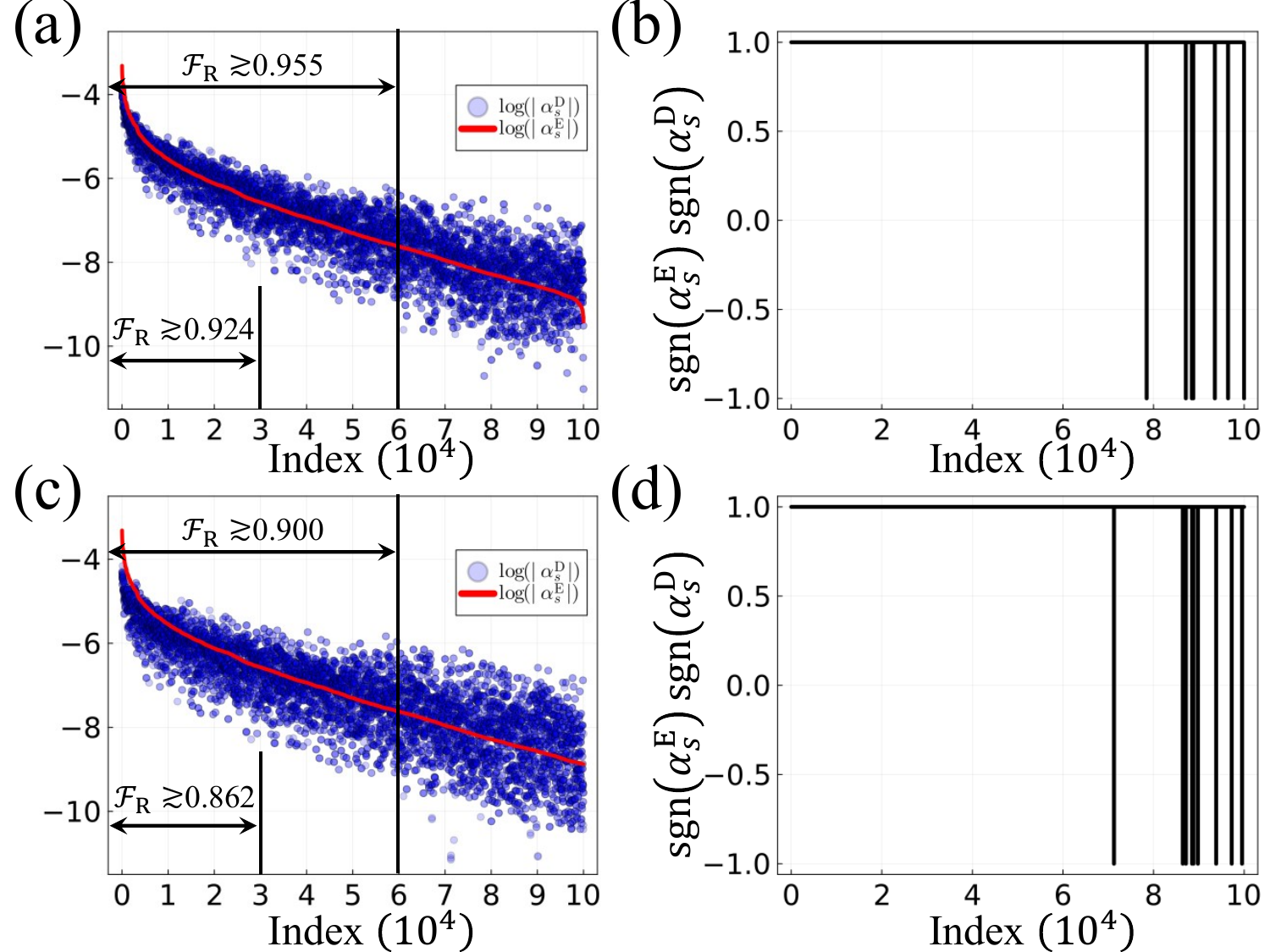}
\caption{Comparison of (a,c) the amplitudes and (b,d) signs between expansion coefficients $\alpha_s^{\text{E}}$ and $\alpha_s^{\text{D}}$ for a $2/3$ hole-doped stripe in the $t$-$J_z$ model with $J=1$ (a,b) and $J=0.6$ (c,d).
The other parameters are  $M=2048$, $L_x=11$, and $L_y=6$.
}\label{figS4}
\end{figure}

\subsection{Fidelity of QCSM}
Because the QCSM~\eqref{eq:effHam} is bottom-up derived from the $t$-$J_z$ model, we can compare its ground-state wavefunction $\ket{\psi_\text{E}}$ with that calculated using DMRG $\ket{\psi_\text{D}}$.
However, this comparison is not straightforward, as the two wavefunctions reside in different Hilbert spaces.
As noted in the main text, we first need to obtain a wavefunction $\ket{\psi^\Omega_\text{D}}$ by projecting $\ket{\psi_\text{D}}$ onto the truncated Hilbert space $\Omega$, and the expansion coefficients are given by $\alpha^\text{D}_s = \braket{s | \psi_\text{D}}$.
For $L_y=6$ and $|\Gamma^z| \leq 5$ in the effective model~\eqref{eq:effHam}, $\Omega$ is spanned by $286,098$ bases, arranged in descending order of the amplitude $\vert \alpha_s^\text{E} \vert$.
Notably, since the model Hamiltonians are always real, the ground-state wavefunction may carry an additional overall sign in either the ED or DMRG calculations.
To facilitate the comparison of the signs of the expansion coefficients $\alpha_s^\text{E}$ and $\alpha_s^\text{D}$, we always fix the expansion coefficients for two-spinon bases with a specified chirality sequence $\chi_1\chi_2 = \Downarrow \Uparrow$ to be positive.

In Fig.~\ref{figS3}, we compare both the amplitudes and signs of the expansion coefficients.
In the smaller Hilbert space $\Omega^<_1$, which includes the first $60,000$ bases, it is very impressive that the sign of $\alpha_s^\text{E}$ and $\alpha_s^\text{D}$ are the same.
This consistency is significant, as the signs often play a critical role in forming the long-range $d$-wave pairing patterns, as stated in the main text.
Meanwhile, the discrepancy between the renormalized fidelity values in $\Omega^<_1$ and $\Omega$ is almost negligible.
Even when the Hilbert space is reduced by a factor of $10$, to $\Omega^<_2$ with $30,000$ bases, the residual fidelity remains high.
This suggests that the effective theory provides an excellent description of the original model, a result of the bottom-up deduction based on a quantum string with a $\pi$-phase shift.
Furthermore, we can notice that the amplitudes of the DMRG-calculated coefficients fluctuate around those obtained using QCSM.
These fluctuations becomes more pronounced at smaller $J$, due to the weakening of the AF background.

\begin{table}[b]
    \centering
    \begin{tabular}{|c|c|c|c|}
	\hline
    	\multicolumn{4}{|c|}{A $2/3$ hole-filled stripe with $L_y=6$}\\
	\hline
        \multirow{2}*{$\alpha$} & \multicolumn{3}{c|}{$M$} \\
        \cline{2-4} & $2,048$ & $4,096$ & $8,192$ \\
	\hline
    	$0$ & $5.55\times10^{-6}$ & $6.18\times10^{-7}$ & $4.95\times10^{-8}$ \\
	\hline
    	$0.2$ & $6.11\times10^{-6}$ & $8.00\times10^{-7}$ & $7.86\times10^{-8}$ \\
	\hline
    	$0.4$ & $1.25\times10^{-5}$ & $2.16\times10^{-6}$ & $2.93\times10^{-7}$ \\
	\hline
    	$0.6$ & $2.81\times10^{-5}$ & $6.37\times10^{-6}$ & $1.15\times10^{-6}$ \\
	\hline
    	$0.8$ & $6.36\times10^{-5}$ & $1.96\times10^{-5}$ & $5.07\times10^{-6}$ \\
	\hline
    	$1$ & $2.31\times10^{-4}$ & $9.12\times10^{-5}$ & $3.15\times10^{-5}$ \\
	\hline
    \end{tabular}\\
    \begin{tabular}{|c|c|c|c|}
    	\hline
    	\multicolumn{4}{|c|}{A $1/2$ hole-filled stripe with $L_y=8$} \\
	\hline
        \multirow{2}*{$\alpha$} & \multicolumn{3}{c|}{$M$} \\
        \cline{2-4}
	& $2,048$ & $4,096$ & $8,192$ \\
	\hline
    	$0$ & $3.35\times10^{-5}$ & $7.71\times10^{-6}$ & $1.35\times10^{-6}$ \\
	\hline
    	$0.2$ & $3.48\times10^{-5}$ & $9.08\times10^{-6}$ & $1.90\times10^{-6}$ \\
	\hline
    	$0.4$ & $6.38\times10^{-5}$ & $2.04\times10^{-5}$ & $5.52\times10^{-6}$ \\
	\hline
    	$0.6$ & $1.15\times10^{-4}$ & $4.52\times10^{-5}$ & $1.51\times10^{-5}$ \\
	\hline
    	$0.8$ & $1.74\times10^{-4}$ & $1.26\times10^{-4}$ & $5.33\times10^{-5}$ \\
	\hline
    	$1$ & $4.12\times10^{-4}$ & $2.95\times10^{-4}$ & $1.96\times10^{-4}$ \\
	\hline
    \end{tabular}
    \begin{tabular}{|c|c|c|c|}
    	\hline
    	\multicolumn{4}{|c|}{A $3/4$ hole-filled stripe with $L_y=8$} \\
	\hline
        \multirow{2}*{$\alpha$} & \multicolumn{3}{c|}{$M$} \\
        \cline{2-4}
	& $2,048$ & $4,096$ & $8,192$ \\
	\hline
    	$0$ & $2.66\times10^{-4}$ & $1.23\times10^{-4}$ & $4.62\times10^{-5}$ \\
	\hline
    	$0.2$ & $2.77\times10^{-4}$ & $1.18\times10^{-4}$ & $4.40\times10^{-5}$ \\
	\hline
    	$0.4$ & $3.16\times10^{-4}$ & $1.43\times10^{-4}$ & $6.07\times10^{-5}$ \\
	\hline
    	$0.6$ & $3.59\times10^{-4}$ & $1.89\times10^{-4}$ & $9.44\times10^{-5}$ \\
	\hline
    	$0.8$ & $4.57\times10^{-4}$ & $2.70\times10^{-4}$ & $1.56\times10^{-4}$ \\
	\hline
    	$1$ & $6.10\times10^{-4}$ & $4.62\times10^{-4}$ & $3.35\times10^{-4}$ \\
	\hline
    \end{tabular}
    \caption{The truncation errors in DMRG at $J_z=0.6$ and $L_x=11$.}
    \label{errors}
\end{table}

\subsection{Truncation errors in the DMRG calculations}
In Table~\ref{errors}, we show the truncation errors at different values of hole filling in the stripe, $L_y$ and bond dimension $M$, and find that $M=8,192$ can ensure truncation errors $\lesssim 3 \times10^{-4}$.

\end{document}